\documentclass[
reprint,
superscriptaddress,
 amsmath,amssymb,
 aps,
prb,
]{revtex4-2}

\usepackage{graphicx}
\usepackage{dcolumn}
\usepackage{bm}
\usepackage{hyperref}
\usepackage[mathlines]{lineno}
\usepackage{float}
\usepackage{xcolor}
\usepackage{upgreek}

\begin{document}


\title{Broad-Wavevector Spin Pumping of Flat-Band Magnons}

\author{Jinlong Wang}
\thanks{These authors contributed equally to this work.}
\affiliation{%
Fert Beijing Institute, MIIT Key Laboratory of Spintronics, School of Integrated Circuit Science and Engineering, Beihang University, Beijing 100191, China.
}
\author{Hanchen Wang}
\thanks{These authors contributed equally to this work.}

\email{hanchen.wang@mat.ethz.ch}

\affiliation{%
Department of Materials, ETH Zurich, Zurich 8093, Switzerland.
}%

\author{Jilei Chen}
\altaffiliation{These authors contributed equally to this work.}
\affiliation{%
International Quantum Academy, Shenzhen 518055, China.
}
\affiliation{%
Shenzhen Institute for Quantum Science and Engineering (SIQSE), and Department of Physics, Southern University of Science and Technology (SUSTech), Shenzhen 518055, China.
}
\author{William Legrand}%
\altaffiliation{These authors contributed equally to this work.}
\affiliation{%
Department of Materials, ETH Zurich, Zurich 8093, Switzerland.
}%
\author{Peng Chen}
\altaffiliation{These authors contributed equally to this work.}
\affiliation{%
Beijing National Laboratory for Condensed Matter Physics, Institute of Physics, University of Chinese Academy of Sciences, Chinese Academy of Sciences, Beijing 100190, China.
}%
\author{\\Lutong Sheng}
\affiliation{%
Fert Beijing Institute, MIIT Key Laboratory of Spintronics, School of Integrated Circuit Science and Engineering, Beihang University, Beijing 100191, China.
}
\author{Jihao Xia}
\affiliation{%
Beijing National Laboratory for Condensed Matter Physics, Institute of Physics, University of Chinese Academy of Sciences, Chinese Academy of Sciences, Beijing 100190, China.
}%
\author{Guibin Lan}
\affiliation{%
Beijing National Laboratory for Condensed Matter Physics, Institute of Physics, University of Chinese Academy of Sciences, Chinese Academy of Sciences, Beijing 100190, China.
}%
\author{Yuelin Zhang}
\affiliation{%
Fert Beijing Institute, MIIT Key Laboratory of Spintronics, School of Integrated Circuit Science and Engineering, Beihang University, Beijing 100191, China.
}%
\author{Rundong Yuan}
\affiliation{%
Fert Beijing Institute, MIIT Key Laboratory of Spintronics, School of Integrated Circuit Science and Engineering, Beihang University, Beijing 100191, China.
}%
\author{Jing Dong}
\affiliation{%
Beijing National Laboratory for Condensed Matter Physics, Institute of Physics, University of Chinese Academy of Sciences, Chinese Academy of Sciences, Beijing 100190, China.
}%
\author{\\Xiufeng Han}
\affiliation{%
Beijing National Laboratory for Condensed Matter Physics, Institute of Physics, University of Chinese Academy of Sciences, Chinese Academy of Sciences, Beijing 100190, China.
}%
\author{Jean-Philippe Ansermet}
\affiliation{%
Institute of Physics, \'Ecole Polytechnique F\'ed\'erale de Lausanne (EPFL), 1015, Lausanne, Switzerland
}
\author{Haiming Yu}
\email{haiming.yu@buaa.edu.cn}
\affiliation{%
Fert Beijing Institute, MIIT Key Laboratory of Spintronics, School of Integrated Circuit Science and Engineering, Beihang University, Beijing 100191, China.
}%
\affiliation{%
International Quantum Academy, Shenzhen 518055, China.
}

\begin{abstract}
We report the experimental observation of large spin pumping signals in YIG/Pt system driven by broad-wavevector spin-wave spin current. 280 nm-wide microwave inductive antennas offer broad-wavevector excitation which, in combination with quasi-flatband of YIG, allows a large number of magnons to participate in spin pumping at a given frequency. Through comparison with ferromagnetic resonance spin pumping, we attribute the enhancement of the spin current to the multichromatic magnons. The high efficiency of spin current generation enables us to uncover nontrivial propagating properties in ultra-low power regions. Additionally, our study achieves the spatially separated detection of magnons, allowing the direct extraction of the decay length. The synergistic combination of the capability of broad-wavevector excitation, enhanced voltage signals, and nonlocal detection provides a new avenue for the electrical exploration of spin waves dynamics.

\end{abstract}

\maketitle


\section{\label{sec:level1}INTRODUCTION}

Magnons, the quanta of spin wave excitations in magnetic materials \cite{kruglyak2010magnonics-1,lenk2011building-6,chumak2015magnon-3,pirro2021advances-2,yu2021magnetic-4,chen2021unidirectional-5}, exhibit remarkable characteristics, including the ability to propagate over centimeter distances without experiencing Joule heating dissipation \cite{serga2010yig-7,chumak2014magnon-8,chen2021reconfigurable-9}. This propagation length significantly surpasses the typical spin diffusion length by several orders of magnitude \cite{bass2007spin-spindiffusion,jansen2012silicon-spindiffusion}. The magnon generation, propagation, and detection scheme can play a role in developing efficient magnon spintronic devices, for example, magnon-based logic gates \cite{chumak2015magnon-3,ganzhorn2016magnon}. The combination of two physical effects: spin pumping (SP) and the inverse spin Hall effect (ISHE) offers a viable approach. Spin pumping (SP) is the process of generating spin currents $\mathbf{J}_\text{s}$ through the excitation of magnons within a ferromagnetic (FM) layer, followed by their injection into an adjacent nonmagnetic (NM) layer \cite{monod1972giant-10,janossy1976spin-11,tserkovnyak2002spin-12}. Subsequently, the inverse spin Hall effect (ISHE) converts these spin currents into charge currents $\mathbf{J}_\text{c}$, more commonly manifesting as voltages \cite{chen2019incoherent-17,mosendz2010detection-18,wei2014spin-back4}. Extensive experimental and theoretical efforts have been devoted to understanding the insightful physical mechanisms related to the SP. This includes the investigation of low damping magnetic material systems such as metallic magnets \cite{ando2008angular-40,gupta2017important-otani2,wang2020spin-back2}, organic-based magnets \cite{liu2020spin-47}, insulating magnets, for example, yttrium iron garnet (YIG) \cite{kajiwara2010transmission-15}, and, more recently, antiferromagnetic hematite \cite{wang2023long,hamdi2023spin,el2023antiferromagnetic}. Studies have explored the relationship between spin pumping and the thickness of FM and NM materials, as well as the impact of excitation frequency and power \cite{haertinger2015spin-back1,castel2012frequency,jungfleisch2015thickness}. Furthermore, there is an emphasis on enhancing interface quality to further improve the spin-mixing conductance \cite{wang2013large-21,takizawa2016spin-otani1,weiler2013experimental-althammer1}. 

The leading process for SP is the generation of $\mathbf{J}_\text{s}$, where the magnitude of $\mathbf{J}_\text{s}$ is proportional to the total number of magnons for various wavevectors $\mathbf{J}_\text{s}$ = $\hbar \sum\limits_{\mathbf{k}} \mathbf{v}_{\mathbf{k}} n_{\mathbf{k}}$ \cite{kajiwara2010transmission-15,maekawa2017spin,maekawa2023spin}. Here, $\mathbf{v}_{k}=\partial \omega_{k} / \partial \mathbf{k}$ is the spin waves velocity and $n_{\mathbf{k}}$ is the number of magnons. The uniform ferromagnetic resonance (FMR) \cite{mosendz2010quantifying-20,wang2013large-21,sun2013damping-22,costache2006electrical-23}, standing spin waves \cite{chen2019incoherent-17,ando2009electric-24,wang2023spin-25}, dipolar spin waves \cite{chumak2012direct-32,PhysRevB.85.094416,sandweg2010enhancement-44, dushenko2015ferromagnetic-45, iguchi2013spin-46}, exchange spin waves \cite{bracher2017detection-26}, and parametrically excited spin waves \cite{manuilov2015spin-27,fukami2016wave-28,ando2012spin-29,sandweg2011spin-30,kurebayashi2011spin-31} are all employed in the generation of $\mathbf{J}_\text{s}$. The majority of previous research has concentrated on the magnons excited with a narrow range of wavevector. This method yields relatively small spin currents $\mathbf{J}_\text{s}$, which, when converted to ISHE voltages, are typically in the range of sub-microvolts or even less at the power level of several milliwatts. This also limits the efficiency of the detection of spin waves, which has consistently garnered significant attention because of its effective information transfer capabilities \cite{chumak2015magnon-3, csaba2017perspectives-logic}.

In this study, we report a systematic investigation of spin pumping signals that arise in the YIG/Pt system, in which a nanometer-scale inductive antenna allows the excitation of magnons with a broad range of wavevectors. By varying the angle between the magnetic field and the wavevector direction, we are able to manipulate the dispersion of the magnons to quasi-flatband. The coexistence of these two conditions leads to a large number of magnons participating in the spin-pumping process, resulting in a significant enhancement of ISHE voltages. Furthermore, we find the nearly 100\% nonreciprocity of nonlocal ISHE voltage in the configuration close to the case where the angle between the magnetic field and the wave vector is from 60° to 80°. The magnon decay length is experimentally estimated by changing the propagation distances.

\section{\label{sec:level1}RESULTS AND DISCUSSION}

\begin{figure}
\centering
\includegraphics[width=86mm]{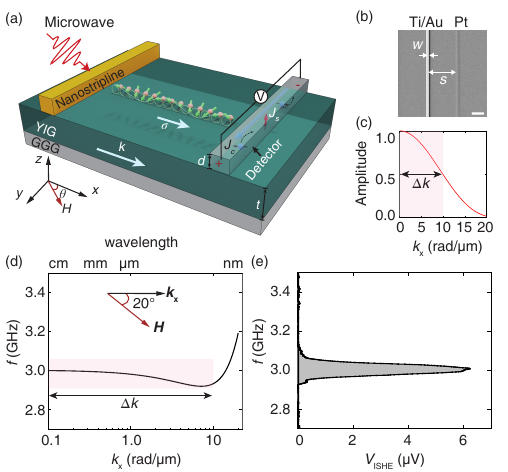}
\caption{(a) Schematic illustration of the broad-wavevector SP set-up. Magnons in the YIG layer are excited using a microwave flowing in a nanostripline antenna. A nano-voltmeter is connected to the Pt bar to detect voltage induced by the charge currents $\mathbf{J}_\text{c}$. $\uptheta$ denotes the in-plane angle between the wavevector $k$ and the applied external field $\boldsymbol{\mathit{H}}$. (b) The scanning electron microscope (SEM) image of the antennas, where the scale bar is 1 $\mu$m. (c) Broad-wavevector distribution calculated by the width of NSL using FFT. The half value of the maximum wavevector excitation, represented by $\Delta{k}$, is chosen to an effective range of wavevector. (d) The magnon dispersion and the mechanism of the broad-wavevector spin pumping process at flat-band scenario. The pink shadow is the effective range of broad-wavevector excitation. (e) The propagating magnons detected by ISHE at the power of $P_\text{MW}$ = 63 $\mu$W (-12 dBm) and $\uptheta$ = 20° with $\mu_\text{0}H$ = 51 mT.}
\label{fig1}
\end{figure}

The sample used in the present study comprises YIG thin films with thickness $t$ = 80 nm grown on (111)-oriented gadolinium gallium garnet (GGG) substrates through magnetron sputtering \cite{liu2014ferromagnetic-36}. The post-annealing process is performed in an oxygen atmosphere at temperatures ranging from 800 to 900 $^{\circ}$C. The Gilbert damping parameter is determined by flipchip ferromagnetic resonance measurements and found to be $\alpha$ = 5.6 $\pm$ 0.2 $\times$ 10$^{-4}$, and the long-range inhomogeneity-caused linewidth is about $\mu_\text{0}\Delta H_\text{0}$ = 0.42 mT, as shown in Fig. S1 in the Supplemental Material \cite{SPM}, consistent with previously reported values \cite{d2013inverse-33,sun2012growth-37}. The Ti/Au for microwave (MW) antenna is deposited by electron beam evaporation and heavy metal Pt (7 nm) for detectors is dc-sputter-deposited at room temperature under a working pressure of 2.8 mTorr and a base pressure of 2.0 $\times$ 10$^{-7}$ Torr. 
\par
Figure~\ref{fig1}$\,$(a) depicts the schematic illustration of the conversion mechanism from propagating magnons to charge current. A nanostripline (NSL) antenna is fabricated on the YIG, and the antenna is connected to a vector network analyzer (VNA) via microwave probes. The injected microwave current generates a time-varying oscillating magnetic field ($\mu_\text{0}h_\text{rf}$) that excites magnons and then propagates $s$ = 2 $\mu$m to the Pt detector. Subsequently, spin-wave spin current is converted into a charge current locally at the detection part. The Pt bar is connected to a nano-voltmeter to detect ISHE voltage ($V_\text{ISHE}$). The magnon dispersion of an 80 nm-thick YIG film, calculated by the dipolar-exchange spin waves theory \cite{kalinikos1986theory-51}, is presented in Fig.~\ref{fig1}$\,$(d). We find that when a magnetic field $\mu_\text{0}{H}$ is applied at a 20° angle with respect to wavevector, a near quasi-flat-band of magnons \cite{gallardo2019flat-FLAT1,chen2022magic-FLAT2,wang2023observation-flat3,tacchi2023experimental-flat4} appears within the wavevector range from 0 to 10 rad/$\mu$m. The  280-nm-wide NSL (Fig.~\ref{fig1}$\,$(b)) offers the ability of the broad range of wavevector excitation. Broad-wavevector distribution calculated by the width of NSL using Fast Fourier transform (FFT)  is shown in Fig.~\ref{fig1}$\,$(c). The half-value of the maximum wavevector excitation is chosen to be an effective width of wavevector, represented by $\Delta{k}$ = 10 rad/$\mu$m, which is enough to cover the wavevector associated with the flat-band magnons. As the excitation frequency approaches a specific value close to the flat-band frequency, a significant population of magnons with various wavevectors get excited and participate in the process of generating spin current after propagating to the detection part. The DC component of the spin current for each wavevector at the interface can be described as, \cite{mosendz2010detection-18,althammer2018pure-althammer2,yang2018fmr-fengyuanyang1}
\begin{equation}
\label{Js}
\mathbf{J}^{k}_\text{s}=\frac{\hbar f}{2} \operatorname{Re}\left(g_{\uparrow \downarrow}\right) P \sin ^{2} \phi_{k},
\end{equation}
where $f$ is the excitation frequency, $\operatorname{Re}\left(g_{\uparrow \downarrow}\right)$ is the real part of the interfacial spin mixing conductance $g_{\uparrow \downarrow}$, $\phi_{k}$ is the precession cone angle of spin waves with wave vector $k$, and $P$ is a factor arising from the ellipticity of the magnetization precession. Considering the number of magnons $n_k \propto \phi^2 _k \propto \mathbf{J}^{k}_\text{s} $ \cite{holstein1940field}, the total spin current generated by magnons with different wavevectors $k$ can be represented as $\mathbf{J}^\text{tot}_\text{s}=\sum\limits_{\mathbf{k}}\mathbf{J}^{k}_\text{s}$. Because of ISHE, the spin current can be converted into charge current  $\mathbf{J}_\text{c}=\frac{2e}{\hbar} \Theta_\text{SH} [\mathbf{J}^\text{tot}_\text{s} \times \sigma],$ and is detected as a voltage via Pt bar. Here, $\Theta_\text{SH}$ represents the spin Hall angle, and $\sigma$ is the spin polarization direction aligned with the magnetization $\boldsymbol{\mathit{M}}$. Finally, the ISHE-induced charge current results in charge accumulation at the two ends of the detector, which can be detected by a nano-voltmeter as the ISHE voltage \cite{mosendz2010detection-18,wang2014scaling-58,czeschka2011scaling-59,shikoh2013spin-60}: 
\begin{equation} 
V_\text{ISHE}=\frac{2e \Theta_\text{SH}}{\hbar}\frac{1}{\sigma_\text{N}d_\text{N}}\lambda_\text{SD}\text{tanh}\left(\frac{d_\text{N}}{{2\lambda}_\text{SD}}\right)L\mathbf{J}^\text{tot}_\text{s},
\label{VISHE_2}
\end{equation}  
where the electron charge $e$, charge conductivities and the thickness of the normal metal layer $\sigma_\text{N}$ and $d_\text{N}$, the spin-diffusion length $\lambda_\text{SD}$, and the effective length $L$ of spin pumping. Fig.~\ref{fig1}$\,$(e) shows the $V_\text{ISHE}$ spectrum of propagating magnons recorded at a distance $s$ = 2 $\mu$m, $\mu_\text{0}{H}$= 51 mT with microwave power $P_\text{MW}$ = 63 $\mu$W. We rule out the possibility of the voltage signal contribution from the spin rectification effect imposed by spin Hall magnetoresistance \cite{harder2016electrical-canminghu1,he2022spin-canminghu2} or spin Seebeck effect \cite{uchida2008observation} (see Fig. S2 in Supplementary Material \cite{SPM}). Under the simultaneous fulfilment of a broad-wavevector excitation and flat-band magnon dispersion, the increased population of magnons leads to the large $V_\text{ISHE}$.

\begin{figure}
\centering
\includegraphics[width=86mm]{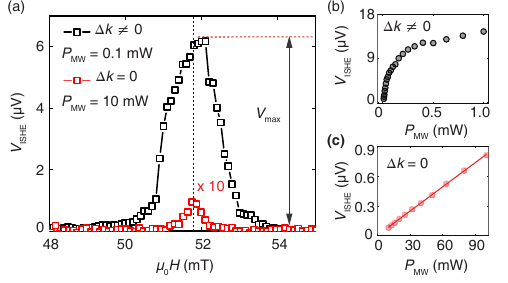}
\caption{(a) $V_{\mathrm{ISHE}}$ as a function of the field at $f$ = 3 GHz for  broad-wavevector spin pumping ($\Delta$k $\ne$ 0 rad/$\mu$m) and uniform spin pumping ($\Delta k = 0$ rad/$\mu$m) with excitation powers of 63 $\mu$W and 10 mW, respectively. $V_\text{max}$ denotes the maximum voltage. (b) Excitation power dependence of maximum value in voltage spectra for $\Delta$k $\ne$ 0 rad/$\mu$m and (c) for $\Delta k = 0$ rad/$\mu$m, respectively. The red solid line is the linear fitting result.}
\label{fig2}
\end{figure}

To attribute the multichromatic magnons to the enhancement of the spin current, we conduct $V_{\mathrm{ISHE}}$  measurements using uniform spin pumping ($V_{\Delta k\ =\ 0}$) (see setup in Fig. S3 \cite{SPM}) for comparison with broad-wavevector spin pumping ($V_{\Delta k\ \ne\ 0}$). The excitation power is 10 mW for $\Delta k$ = 0 measurements and 63 $\mu$W for $\Delta k$ $\ne$ 0 measurements, respectively,  maintaining a linear precession region (see Figs.~\ref{fig2}$\,$(b) and 2(c)). We fix $f$ = 3 GHz and sweep the magnetic field. The $V_\text{ISHE}$ curves are depicted in Fig.~\ref{fig2}$\,$(a), where the maximum voltage $V_\text{max}$ for $V_{\Delta k\ \ne\ 0}$ is 6.2 $\mu$V and the black dot line denotes the ferromagnetic resonance field $\mu_\text{0}H_\text{FMR}$. The maximum value for $V_{\Delta k\ =\ 0}$ at $\mu_\text{0}H_\text{FMR}$ is 0.09 $\mu$V. Compared to the measurements with $\Delta k$ = 0, the voltage observed in the $\Delta k\ne$ 0 measurements shows a significant enhancement. In Eq.~\ref{VISHE_2}, one can find that there are many parameters related to the $V_\text{ISHE}$. The $\lambda_\text{SD}$ $\approx$ 8 nm for Pt , surpasses $d_\text{N}$, where  $d_\text{N}$ is 5 nm for $\Delta k=$ 0 and 7 nm for $\Delta k\ne$ 0 measurements. Considering that the term $\frac{\lambda_\text{SD}}{d_\text{N}}\text{tanh}(\frac{d_\text{N}}{2\lambda_\text{SD}})$ is essentially constant for $\lambda_\text{SD}$ $> d_\text{N}$ due to the limitation of film thickness \cite{wang2014scaling-58}, it has a feeble effect on the measured voltage. Consequently, we have $\frac{V_\text{$\Delta k\ne$ 0}}{V_\text{$\Delta k=$ 0}}=\frac{L_\text{$\Delta k \ne$ 0}}{L_\text{$\Delta k =$ 0}}\frac{\sum\limits_{\mathbf{k}}\mathbf{J}^{k}_\text{s}}{\mathbf{J}^{0}_\text{s}}$, where $\mathbf{J}^{0}_\text{s}$ is the spin current generated by uniform spin pumping. One needs a factor $\delta$ meet $\sum\limits_{\mathbf{k}}\mathbf{J}^{k}_\text{s}$ = $\delta\ \mathbf{J}^{k}_\text{s}$, by which the whole contribution of the broad-wavevector magnons may be qualitatively extracted from the measured value of the spin-pumping voltage. According to Eq.~\ref{Js}, $\mathbf{J}_\text{s}$ is the function of the cone angle $\phi$. $\phi$ is generally small (1° or less) and can be expressed as sin$\phi_\text{k} \approx \phi_\text{k}$ = $\frac{\gamma \mu_\text{0}h_\text{rf}}{4\pi\alpha^{k}_{\rm eff} f}$, where $\alpha^{k}_{\rm eff}$ is the wave vector-dependent effective Gilbert damping. Combining it with $\delta$, we have:
\begin{equation} 
\delta=\frac{L_\text{$\Delta k \ne $ 0}}{L_\text{$\Delta k = $ 0}}{\left(\frac{\mu_\text{0}h_\text{rf}^\text{$\Delta k \ne $ 0}}{\mu_\text{0}h_\text{rf}^\text{$\Delta k = $ 0}}\right)}^2{\left(\frac{\alpha_\text{YIG/Pt}}{\alpha_\text{eff}^k}\right)}^2.
\label{VISHE_3}
\end{equation}
\par
We normalize the parameters spin pumping effective length $L$, $\mu_\text{0}h_\text{rf}$, and effective damping $\alpha$ for $\Delta k \ne 0$ and $\Delta k = 0$ measurements. Assuming that the broad-wavevector magnons are uniformly excited along the length of the NSL, we can approximate the length of the Pt detector as $L_{\Delta k \ne 0}$, which is approximately 100 $\mu$m. Considering a 120 nm-thin NSL of width $\omega$ with its centre at $x$ = $z$ = 0, the electromagnetic field distribution is calculated, and it reveals the maximum $\mu_\text{0}h_\text{rf}$ value of 0.93 mT at the central position with input $P_\text{MW}\ =\ $63 $\mu$W (see Fig.~S4(a) \cite{SPM}). The $L(\mu_\text{0}h_\text{rf})^2$ for $\Delta k \ne $ 0 is estimated to 0.09 mT$^2$mm. The value $L(\mu_\text{0}h_\text{rf})^2$ = 0.13 mT$^2$mm is calculated for $\Delta k = $ 0 by integrating over the effective region influenced by the antenna's field when $P_\text{MW}\ =\ $10 $m$W  (see Fig.~S4(b) \cite{SPM}). From Fig.~\ref{fig2}$\,$(a), for $V_{\Delta k\ \ne\ 0}$, the magnetic field $\mu_\text{0}H$ corresponding to the $V_\text{max}$ is 52.2 mT. The frequency $f$ of spin-wave with wavevector $k$ \cite{kalinikos1986theory-51,yu2014magnetic-52,wang2022hybridized-53,chen2019excitation-54}:

\begin{figure*}
\centering
\includegraphics[width=150mm]{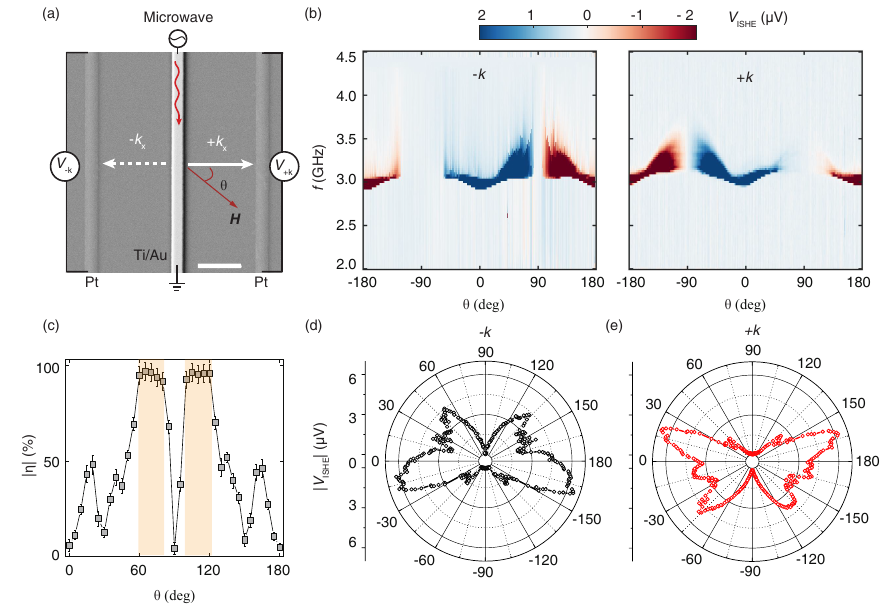}
\caption{(a) The device wherein two identical Pt strips are placed symmetrically in the left ($\text V_{-k}$) and right ($\text V_{+k}$) sides of NSL with same propagation distance 2 $\mu$m, where scale bar is 1 $\mu$m. (b) Angle-dependent spectra of $V_\text{ISHE}$ for magnons with opposite propagating directions at $\mu_\text{0}H$ = 52 mT and $P_\text{MW}$ = 63 $\mu$W. (c) $\uptheta$ dependence of $V_\text{ISHE}$ nonreciprocity $\eta$. (d) Angular dependence of the absolute voltage with the extraction of maximum $V_\text{ISHE}$ values at each angle from -$k$ spectrum and (e) from +$k$ spectrum.}
\label{fig3}.
\end{figure*}

\begin{widetext}  
\begin{equation} 
f=\frac{|\gamma|\mu_0}{2\pi}\left[\left(H+\frac{2A}{\mu_0 M_{\rm s}}k^{\rm 2}\right)\times\left(H+\frac{2A}{\mu_0 M_{\rm s}}k^{\rm 2}+FM_{\rm s}\right)\right]^\frac{1}{2},
\label{dispersion1}
\end{equation}  
\begin{equation}
F=1-\left( 1-\frac{1-e^{-kt}}{kt}\right) \rm{cos}^{\rm 2}\uptheta\\+\it{\left(\frac{M_{\rm s}}{H+\frac{{\rm 2}A}{\mu_{\rm 0} M_{\rm s}} k^{\rm 2}}\right)\left(\frac{{\rm 1}-e^{-{\rm 2}kt}}{{\rm 4}}\right)}\rm{sin}^2 \uptheta,
\label{Fin_dispersion}
\end{equation} 
\end{widetext}
\noindent where $|\gamma|=2\pi\times$28 GHz/T is the gyromagnetic ratio, $A$ = 3 pJ/m is the exchange stiffness constant\cite{chen2019excitation-54}, $t$ is the thickness of the YIG layer, and $F$ is the dipolar array factor. When excitation frequency $f$ = 3 GHz, the maximum $k$ corresponding to 52.2 mT is 2.5 rad/$\mu$m (see Fig. S6 \cite{SPM}). Given the varying damping between high-$k$ magnons and $k$ = 0 magnons, which tends to increase with increasing the wavevector, we derived the effective damping $\alpha_{\rm eff}$  from the wavevector-dependent relaxation time $\tau(k)=[2\pi\alpha_\text{eff}f(k)]^{-1}$ \cite{wang2022hybridized-53,prabhakar2009spin}. The effective damping $\alpha^k_\text{eff}$ is approximately 6.5$\times$10$^{\rm -4}$ when $k$ is 2.5 rad/$\mu$m. The damping of YIG/Pt bilayer $\alpha_\text{YIG/Pt}$ for $\Delta k = $ 0 measurements is 7.8 $\times$ 10$^{-4}$ (see Fig. S5 \cite{SPM}).~Taking into account the combined influence of $\frac{L_\text{$\Delta k \ne $ 0}}{L_\text{$\Delta k = $ 0}}{\left(\frac{h_\text{rf}^\text{$\Delta k \ne $ 0}}{h_\text{rf}^\text{$\Delta k = $ 0}}\right)}^2{\left(\frac{\alpha_\text{YIG/Pt}}{\alpha_\text{eff}^k}\right)}^2$, it gives a ratio nearly 1, compared with the ratio of the maximum voltage observed in the experiment $\frac{\text V_\text{$\Delta k \ne $ 0}}{\text V_\text{$\Delta k = $ 0}}$ = 60, there should have a factor of 60 from $\delta$. The evolution of the $V_\text{ISHE}$ signals as a function of wavevector for excitations also demonstrates the effectiveness of broad-wavevector spin pumping (see Fig. S7 \cite{SPM}). Figs.~\ref{fig2}$\,$(b) and 2(c) summarize the maximum value in a voltage spectrum under different excitation powers for $\Delta k \ne $ 0 and $\Delta k\ =\ 0$, respectively. The spin pumping with $\Delta k\ =\ 0$ exhibits a linear relationship with $P_\text{MW}$ up to the maximum microwave power of 100 mW (Fig.~\ref{fig2}$\,$(c)). In comparison, the signal of spin pumping with $\Delta k \ne $ 0 maintains in the linear region only up to 0.1 mW. This demonstrates that the spin pumping with $\Delta k \ne $ 0 requires much lower microwave power than the $\Delta k = 0$ to achieve the same spin current, and also much easier step into the nonlinear regime \cite{sheng2023nonlocal}.

In order to gain more insight into the magnon-driven ISHE, we study its angular dependence  (Fig.~\ref{fig3}$\,$) with a MW power of 63 $\mu$W. First, two identical Pt detectors are positioned on the two opposite sides of the NSL with a separated distance of 2 $\mu$m.~At 52 mT, we record the angle-dependent spectra $V_\text{ISHE}$ for both -$k$ and +$k$ magnons (Fig.~\ref{fig3}$\,$(b)) by rotating the in-plane magnetic field. We extract the $V_\text{max}$ at each angle (Figs.~\ref{fig3}$\,$(d) and 3(e)). The output voltage exhibits an asymmetric dependency resembling a butterfly pattern. At the excitation part, the varying microwave absorption capability of the magnetic material under different $\uptheta$ leads to an angular dependency in the NSL's excitation efficiency, described using $A\text{sin}^2\uptheta + B$ \cite{el2023antiferromagnetic}. At the detection part, the ISHE follows a cos$\uptheta$ law \cite{ando2011inverse-50}. Merely considering the angular dependencies of excitation and detection doesn't align with the experimental outcomes (see Fig. S8 \cite{SPM}). Therefore, we consider that the propagation of magnons contributes to additional asymmetry. As $\uptheta$ gradually rotates from 0° towards 90°, the spin wave mode transitions from a volume mode to a surface mode, exhibiting noticeable nonreciprocal propagating (see Fig. S8 within the Supplemental Material \cite{SPM}). We quantify the $V_\text{ISHE}$ nonreciprocity in terms of  the ratio,
\begin{equation}
    \eta = \frac{V_{-k} - V_{+k}}{V_{-k} + V_{+k}},
\end{equation}
where $|\eta|$ = 100\% indicates perfect unidirectional $V_\text{ISHE}$ detection. We show the $|\eta|$ extracted from the experiments as a function of $\uptheta$ in Fig.~\ref{fig3}$\,$(c), which exhibits nearly 100\% when $\uptheta$ is rotated between 60° and 80°. Due to the finite thickness of YIG, the amplitude distribution of precession profile along the thickness direction is non-uniform, particularly in the context of dipolar-exchange spin waves. Consequently, the Damon-Eshbach (DE) mode has the capability to induce a spin-wave spin current at the top or bottom surfaces of the material. As an illustrative example, when considering the case of $+k$ magnons, apart from the detection restrictions causing $V_\text{ISHE}$ signal of zero within the DE mode, in the vicinity of the DE mode at angles $\uptheta$, roughly between 60° and 80°, a close-to-zero $V_\text{ISHE}$ can also be observed. This phenomenon underscores the interfacial nature of the spin pumping process, implying that the majority of magnons likely propagate along the bottom surface. This, in turn, suggests that the primary source of the spin current is surface-bound magnons, as opposed to magnons traversing the entire thickness of the material.

\begin{figure}
\centering
\includegraphics[width=86mm]{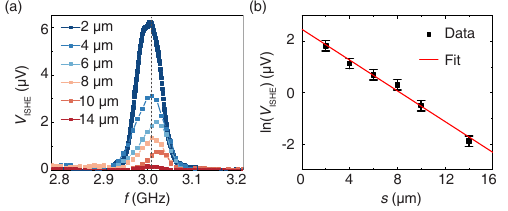}
\caption{(a) Frequency-dependent $V_{\mathrm{ISHE}}$ lineplots measured in devices with different propagation distances at $\mu_\text{0}H$ = 52.2 mT, $\uptheta$ = 20° and $P_\text{MW}$ = 63 $\mu$W. The grey dash line corresponds to frequency $f$ = 3 GHz. (b) The dependence of the logarithm of ln($V_{\mathrm{ISHE}}$) as a function of propagation distance at 3 GHz. The red line is the linear fitting using Eq.~\ref{decay}.}
\label{fig4}
\end{figure}

The spatially separated detection enables us to measure the decay length of magnons. Consequently, keeping the excitation NSL unchanged, we conduct measurements to explore the $V_\text{ISHE}$ dependence on distance. Fig.~\ref{fig4}$\,$(a) show the $V_\text{ISHE}$ recorded at $\mu_\text{0}H$ = 52.2 mT with propagation distances $s$ = 2, 4, 6, 8, 10, and 14 $\mu$m. The longer the propagation distance is, the less magnons can then be detected, and therefore the weaker ISHE signal. With a fixed frequency at 3 GHz, the maximum values of $V_{\mathrm{ISHE}}$ at different propagation distances are extracted. The dependence of the logarithm of ln($V_{\mathrm{ISHE}}$) as a function of $s$ is shown in Fig.~\ref{fig4}$\,$(b) and fitted them by, 
\begin{equation}
    \text{ln}\left(\frac{V_{\mathrm{ISHE}}}{C_\text{m}}\right) = -\frac{s}{\lambda_\text{m}},
\label{decay}
\end{equation}
where $C_\text{m}$ and $\lambda_\text{m}$ are distance-independent prefactor and magnon decay length, respectively. By fitting the results of the experiment, the decay length $\lambda_\text{m}$ is extracted about 3.3 $\mu$m, which compares well to previous experimental results \cite{qin2018propagating, talalaevskij2017magnetic-decay}.  Decay length $\lambda_\text{m}$ = $v_\text{g}$/2$\pi\alpha{f}$, is largest for magnons with high group velocity or small damping. Unlike the conventional uniform excitation SP, which necessitates the deposition of Pt on the entire surface of the YIG, creating a YIG/Pt heterostructure, our approach obviates this requirement, thus mitigating the increase in YIG's damping coefficient. This approach allows us to preserve a relatively large decay length. Moreover, spatially separated detection offers distinct advantages in terms of facilitating gating and manipulating spin waves. When combined with the efficient excitation of magnons, it paves the way for a novel route in the generation, propagation, and detection of spin currents, which is essential for the development of magnon-based logic devices.

\section{\label{sec:level1}CONCLUSION}

In conclusion, we experimentally observe significant voltage signals from broad-wavevector excitation flat-band magnons spin pumping in thin YIG films. The utilization of nanoscale stripline enables efficient angular momentum transfer of flat-band magnons across a broad range of wavevectors into the spin current. The generation of spin-wave spin current with $\Delta k$ = 2.5 rad/$\mu$m is found to be nearly 60 times greater than that of $\Delta k$ = 0 rad/$\mu$m one. We also clarify that spin pumping primarily originates from magnons at the surface rather than spanning the full thickness, supported by the nonreciprocal propagation of magnetostatic surface spin waves. The ability of unidirectional detection of $V_\text{ISHE}$ in combination with long-distance propagating magnons can become a key functionality in reconfigurable nanomagnonic logic and computing devices. Our find allows for the downsizing of input microwave power and spin pumping structures, all while preserving adequately robust signals and the dynamic properties of spin waves. The periodic Dzyaloshinskii-Moriya coupling or moir{\'e} pattern also induced flat-band \cite{gallardo2019flat-FLAT1,chen2022magic-FLAT2,wang2023observation-flat3,tacchi2023experimental-flat4}. In the future, combining this with broad-wavevector spin pumping could offer new insights into the spin waves dynamics regime.

\begin{acknowledgments}
The authors thank G. E. W. Bauer, P. Gambardella, K. Yamamoto, and S. Maekawa for their helpful discussions. We wish to acknowledge the support by the National Key Research and Development Program of China Grant No. 2022YFA1402801, by NSF China under Grants No. 12074026, No. 52225106, and No. U1801661, and by Shenzhen Institute for Quantum Science and Engineering, Southern University of Science and Technology (Grant No. SIQSE202007). H. W. acknowledge support by China Scholarship Council (CSC) under Grant No. 202206020091 and W. L. acknowledge support by an ETH Zurich Postdoctoral Fellowship (21-1 FEL-48).
\end{acknowledgments}

\bibliography{apssamp}

\end{document}